\documentclass[a4paper,11pt]{article}
\usepackage{pos}

\usepackage{amsmath}
\usepackage{braket}

\usepackage{algorithm}
\usepackage{algpseudocode}
\usepackage{amsfonts}

\usepackage{tikz}
\usetikzlibrary{decorations.pathreplacing,calc}
\newcommand{\tikzmark}[1]{\tikz[overlay,remember picture] \node (#1) {};}

\newcommand*{\AddNote}[4]{%
    \begin{tikzpicture}[overlay, remember picture]
        \draw [decoration={brace,amplitude=0.5em},decorate, thick, black]
            ($(#3)!(#1.north)!($(#3)-(0,1)$)$) --  
            ($(#3)!(#2.south)!($(#3)-(0,1)$)$)
                node [align=center, text width=.3\textwidth, pos=0.5, anchor=west] {#4};
    \end{tikzpicture}
}%

\def\SymbReg{\textsuperscript{\textregistered}}
\def\mkl{Intel\SymbReg{} Math Kernel Library (Intel\SymbReg{} MKL)}

\usepackage{subcaption}
\usepackage{tabularx}

\usepackage{enumitem}
\setenumerate[1]{label=(\arabic*)}
\usepackage{lstautogobble}
\title{Performance Optimization of Baryon-block Construction in the Stochastic LapH Method}
\ShortTitle{Baryon-block Construction in the Stochastic LapH Method}

\author*[a,b]{Phuong Nguyen}
\author[a]{Ben H\"{o}rz}

\affiliation[a]{Intel Deutschland GmbH,
Dornacher Str. 1, 85622 Feldkirchen, Germany}

\affiliation[b]{Technical University Munich, Boltzmannstr. 3,  85748 Garching, Germany}

\emailAdd{phuong.nguyen@tum.de}
\emailAdd{ben.hoerz@intel.com}

\abstract{Implementations of measurement kernels in high-level Lattice QCD frameworks enable rapid prototyping, but can leave hardware capabilities significantly underutilized. This is an acceptable tradeoff if the time spent in unoptimized routines is generally small. The computational cost of modern spectroscopy projects however can be comparable to or even exceed the cost of generating gauge configurations and computing solutions of the Dirac equation. One such key kernel in the stochastic LapH method is the computation of baryon blocks; we discuss several implementation strategies and achieve a 7.2x speedup over the current implementation on a system with Intel\SymbReg{} Xeon\SymbReg{} Platinum 8358 processors, formerly Ice Lake.}

\FullConference{%
The 39th International Symposium on Lattice Field Theory,\\
8th-13th August, 2022,\\
Rheinische Friedrich-Wilhelms-Universität Bonn, Bonn, Germany
}

\begin{document}
\maketitle

\section{Introduction}
With the advent of modern spectroscopy methods for multi-hadron systems \cite{HadronSpectrum:2009krc,Morningstar:2011ka,Detmold:2019fbk}, ever more complicated physical systems are coming into reach.
While a variety of two-meson systems including with several coupled channels have been investigated (see \cite{Liu:lat22} for a plenary review at this conference), and more recently even studies of three-meson systems have started appearing~\cite{Horz:2019rrn,Blanton:2019vdk,Mai:2019fba,Culver:2019vvu,Fischer:2020jzp,Hansen:2020otl,Alexandru:2020xqf,Blanton:2021llb} (also subject of a topical review this year~\cite{Fernando:lat22}), the situation for systems with baryons is comparably less advanced.
Hadron interactions involving baryons are however among the fundamental building blocks for an understanding of nuclear physics rooted in QCD (for a review, see for instance~\cite{Drischler:2019xuo}).
Examples include two-nucleon interactions, which -- when determined at sufficiently light pion mass -- can serve as a validation system to establish the reliability of these kinds of lattice QCD calculations~\cite{Francis:2018qch,Green:2021qol,Horz:2020zvv,Amarasinghe:2021lqa}; three-nucleon interactions, which are difficult to access phenomenologically and hence present a great opportunity for lattice QCD to provide useful data; a variety of meson-baryon systems, for instance nucleon-pion scattering in the isospin $I=3/2$ channel featuring the $\Delta(1232)$ resonance~\cite{Andersen:2017una,Silvi:2021uya,Bulava:2022vpq}.

Even though modern spectroscopy methods present no conceptual difficulties generalizing to systems involving baryons, there are a few challenges in practice.
Baryonic systems typically suffer from a worse signal-to-noise ratio than purely mesonic systems, requiring larger amounts of statistics to obtain meaningful results.
In addition to necessitating more correlation-function samples, every individual sample tends to be more computationally expensive compared to the mesonic sector due to the increased number of quark fields.
At the correlator-construction level, algorithms eliminating redundant computations have been devised to alleviate the proliferation of Wick contractions \cite{Doi:2012xd,Detmold:2012eu,Gunther:2013xj,Horz:2019rrn}.

This work is concerned with improving the efficiency of the computation of baryon functions in the stochastic LapH method~\cite{Morningstar:2011ka}.
In the stochastic LapH framework, baryon blocks are rank-three tensors in dilution indices, carrying additional labels identifying the baryon operator (flavor, spin, hadron momentum) as well as the three noises used for the stochastic estimate of the quark propagators in the LapH subspace.
Correlators are then computed through tensor contractions over dilution indices of those baryon functions as governed by Wick's theorem.
This beneficial property of the stochastic LapH method -- the evaluation of complicated multi-hadron correlation functions is reduced to tensor contractions involving blocks representing the constituent hadrons -- enables the re-use of baryon blocks for a wide variety of physical systems.
The optimizations presented in this work are hence immediately applicable to a breadth of calculations involving baryons.

This contribution is organized as follows: \autoref{sec:bblock} defines the baryon-block kernel and discusses its computational characteristics, \autoref{sec:impl} contrasts several implementation strategies, and benchmark results are presented in \autoref{sec:results}.

\section{Baryon blocks in the stochastic LapH method}
\label{sec:bblock}
The stochastic LapH method~\cite{Morningstar:2011ka} is a stochastic variant of distillation~\cite{HadronSpectrum:2009krc} which avoids the $V^2$ scaling with the spatial simulation volume $V$ by stochastically estimating the quark propagator projected into the LapH (or distillation) subspace.
A useful quantity for the computation of multi-hadron correlation functions involving baryons is the (single-site) baryon function defined per time slice of the simulation volume,
\begin{align}
  B_{d_1 d_2 d_3}^{(\vec p, \Lambda, \mu, \eta)} = c^{(\vec p, \Lambda, \mu)}_{\alpha \beta \gamma} \sum_{\vec x} \mathrm{e}^{-\mathrm{i} \, \vec p \vec x} \epsilon_{abc} q^{(\eta_1, d_1)}_{\alpha a \vec x}  q'^{(\eta_2, d_2)}_{\beta b \vec x} q''^{(\eta_3, d_3)}_{\gamma c \vec x}, \label{eqn:bfunc}
\end{align}
with color indices $a$, $b$, $c$, and the summation runs over the sites of a three-dimensional time slice of the lattice.
The relevant combinations of spin indices $\alpha$, $\beta$, $\gamma$ are selected according to the group-theoretical projection coefficients $c^{(\vec p, \Lambda, \mu)}_{\alpha \beta \gamma}$ for a given hadron momentum $\vec p$, irrep $\Lambda$ and irrep row $\mu$~\cite{Morningstar:2013bda}.
The quark fields $q$, $q'$, $q''$, which have been projected into the LapH subspace, are obtained by repeatedly solving the Dirac equation with diluted stochastic sources identified by a noise label $\eta$ and dilution index $d_1 = 1, \dots, N_\mathrm{dil}$.
Crucially, those solutions of the Dirac equation for a given noise need only be computed once, and can then be cheaply stored on disk and used for various multi-hadron projects by reconstructing the smeared quark fields from their coefficients $Q$ in the basis of eigenvectors of the three-dimensional gauge-covariant Laplacian $\phi^{(l)}$, $l=1,\dots,N_\mathrm{ev}$, which is used to define the LapH subspace,
\begin{align}
  q^{(\eta, d)}_{\alpha a \vec x} = \sum_{l=1}^{N_\mathrm{ev}} Q^{(\eta, d)}_{\alpha l} \phi^{(l)}_{a \vec x}, \label{eqn:reconq}
\end{align}
and similarly for $q'$ and $q''$, which differ only in their coefficients $Q$, but with the same eigenvectors $\phi$.

Both the group-theoretical projection involving the spin indices and the bookkeeping of noise indices in \eqref{eqn:bfunc} are handled by the calling application, leaving
\begin{align}
    B_{d_1 d_2 d_3}^{(\vec p)} = \sum_{\vec x} \mathrm{e}^{-\mathrm{i} \, \vec p \vec x} \epsilon_{abc} Q^{(1)}_{d_1 l_1} Q^{(2)}_{d_2 l_2} Q^{(3)}_{d_3 l_3} \phi^{(l_1)}_{a \vec x} \phi^{(l_2)}_{b \vec x} \phi^{(l_3)}_{c \vec x}, \label{eqn:bkern}
\end{align}
where summation over the eigenvector indices $l_1$, $l_2$ and $l_3$ is implied, as the computational kernel, which is called many times for different noise and spin combinations,~i.e.~different quark field coefficients, but with the same momentum set and Laplacian eigenvectors.

Depending on the sizes of $N_\mathrm{ev}$ and $N_\mathrm{dil}$, as well as how many times the kernel \eqref{eqn:bkern} is called per time slice, one of the following two different approaches is preferable.

For moderate values of $N_\mathrm{ev}$, \eqref{eqn:bkern} can be efficiently computed using a two-step procedure: During the setup stage, the mode-triplets
\begin{align}
  T_{l_1 l_2 l_3}^{\vec p} = \sum_{\vec x} \mathrm{e}^{-\mathrm{i} \, \vec p \vec x} \epsilon_{abc} \phi^{(l_1)}_{a \vec x} \phi^{(l_2)}_{b \vec x} \phi^{(l_3)}_{c \vec x}, \label{eqn:mt}
\end{align}
which are spin-, noise- and flavor-blind, are computed and kept in memory.
All lattice-sized objects in \eqref{eqn:bkern} have then been consumed, and the baryon function for a given set of quark-field coefficients can be computed by tensor-contracting them onto the precomputed mode-triplet.
Those tensor contractions can be performed with high performance, so the majority of the runtime tends to be associated with the initial setup phase, which needs to be amortized over many kernel invocations.
The major drawback of this mode-triplet approach is the need to keep one $N_\mathrm{ev}^3$-sized object per momentum in memory%
\footnote{Based on the symmetries of \eqref{eqn:mt}, only ${N_\mathrm{ev} \choose 3}$ elements of a mode-triplet are independent. Exploting that symmetry with a sparse storage scheme however complicates the subsequent tensor contractions of quark-field coefficients onto the mode-triplet.}.

For large number of eigenvectors $N_\mathrm{ev}$, a more economical approach is to first reconstruct the quark fields from the coefficients as per \eqref{eqn:reconq}, and subsequently perform the reduction \eqref{eqn:bkern} over sets of lattice-sized objects.
The quark-field reconstruction can be efficiently implemented using matrix-matrix multiplication and reduces the complexity of subsequent lattice-sized reductions to $N_\mathrm{dil}^3$ (rather than $N_\mathrm{ev}^3$ for the mode-triplet approach), which however must be performed for every kernel invocation.

In view of the requirements of baryon calculations in large volumes -- such as the E250 ensemble~\cite{Mohler:2017wnb} generated by the CLS effort~\cite{Bruno:2014jqa,Bruno:2016plf}, where employing the mode-triplet approach is not feasible%
\footnote{First results in the mesonic sector presented in \cite{Paul:2021pjz} used the analogous mode-doublet approach for meson construction, which is still affordable due to its slightly weaker $N_\mathrm{ev}^2$ scaling. For the $N_\mathrm{ev} = 1536$ employed in that work, the mode triplet on the other hand occupies $1.8 \, \mathrm{TB}$ of memory already for the moderate number of momenta $N_\mathrm{mom} = 33$, clearly making this approach impractical.} -- the goal of this work is to provide an efficient implementation of \eqref{eqn:bkern}, which utilizes the great compute capabilities of modern hardware by exploiting the $N_\mathrm{dil}^3$ compute complexity with only linear-in-$N_\mathrm{dil}$ memory traffic.

\section{Implementation details}
\label{sec:impl}
%---------------------------------------------------------------------------------------------

    \begin{algorithm}[tbh]
        \caption{Cache blocking algorithm with pseudo code}
        {\fontsize{10pt}{10.5pt}\selectfont
        \label{alg:opt}
        \begin{algorithmic}[1]
            \Require{$q_1, q_2, q_3, phase$}
            \Ensure{$baryon$}
            \Statex
            \State{! $Block{D_i} \gets N_{D_i}\, / \, Bsize{D_i}$ \, ($i=1,2,3$)}
            \State{! $BlockX \gets N_{X}\, / \, Bsize{X}$}
            \Function{BaryonConstruct}{$q_1, q_2, q_3, phase$}
            \For{$Block{D_1}$} \textbf{in parallel} \tikzmark{top}
            \For{$Block{D_2}$} \textbf{in parallel}
            \State{\textit{tmpBuf} $\gets 0.$}
            \For{\textbf{each} $Block{X}$} \label{alg:opt:blockx}
            \For{$d_1 \gets 1$ to $Bsize{D_1}$}
            \For{$d_2 \gets 1$ to $Bsize{D_2}$}
            \For{$x \gets 1$ to $Bsize{X}$}
            \State{$diq(d_1, d_2, :, x) \gets q_1(\tilde{d_1}, :, x) \times q_2(\tilde{d_2}, :,x)$} \tikzmark{right}
            \EndFor
            \EndFor
            \EndFor
            \For{\textbf{each} $Block{D_3}$}
            \For{$d_3 \gets 1$ to $Bsize{D_3}$} \tikzmark{bottom}
            \For{\textbf{each} $diq_{i}$ \textbf{in} $diq$}
            \For{$x \gets 1$ to $Bsize{X}$}
            \State{$singlet(d_1, d_2, d_3, x) \gets diq_i \times q_3(\tilde{d_3}, :, x)$\textcolor{white}{space}}
            \EndFor
            \EndFor
            \EndFor
            \State{\textit{tmpBuf(:)} $\gets$ \textit{tmpBuf(:)} + $singlet \times phase(:)$} \Comment{Intel\SymbReg~MKL JIT GEMM}
            \EndFor
            \EndFor
            \State{$baryon \gets$ \textit{tmpBuf}}
            \EndFor \textbf{ parallel}
            \EndFor \textbf{ parallel}
            \State \Return {$baryon$}
            \EndFunction
        \end{algorithmic}
        \AddNote{top}{bottom}{right}{Cache Blocking in $N_{D_1}$, $N_{D_2}$, $N_{D_3}$, $N_X$}}
    \end{algorithm}

The quark-field reconstruction can be performed efficiently using matrix-matrix multiplication, for which highly optimized implementations are available for all hardware architectures.
Hence, in the following section, we focus on optimizing the baryon-block calculation given the reconstructed quark fields $q_1$, $q_2$, $q_3$.

The optimized algorithm is shown in Algorithm~\ref{alg:opt}.
Typically, the number of requested hadron momenta $N_\mathrm{mom}$ is much smaller than the number of allowed momenta (e.g.~$33 \ll 64^3$); therefore using a fast Fourier transform is not beneficial.
Hence, the phase factor $e^{-\mathrm{i} \vec p \vec x}$ for the momentum projection in \eqref{eqn:bkern} can be precomputed and re-used for several kernel invocations.

Cache blocking techniques are employed in conjunction with an appropriate data layout to optimize data locality.
Blocking is implemented both in the spatial indices $x$ and the three dilution indices $d_1, d_2, d_3$.
The blocking in $x$ allows the kernel to exploit the available inherent input reuse.
For example, each block of $q_1$ input can be kept in the cache and re-used for different $diq$ calculations with different $q_2$ since the input size is small enough to stay in the cache (ll.~8-14).
The blocking in $d_1, d_2, d_3$ enables the kernel to keep the intermediate data (\textit{diq}, \textit{singlet}, \textit{tmpBuf}) in cache and use it for subsequent calculations (ll.~16-22).
A suitable data memory layout is (in row-major convention)
    $N_\mathrm{dil} \times N_\mathrm{BlockX} \times N_\mathrm{color} \times N_\mathrm{BsizeX}$
for the input $q_1, q_2, q_3$,
ensuring that the data is accessed contiguously in $x$ for each each color component. 
Furthermore, this data layout stores the BlockX-sized chunks for the three colors of $q_1, q_2, q_3$ adjacently, enhancing spatial locality in the calculations of \textit{diq} and \textit{singlet}.

% (3) JIT
The small matrix-matrix multiplication in l.~23 to
compute $\textit{tmpBuf(:)} += singlet \times phase(:)$,
utilizes the \mkl{} with just-in-time (JIT) code generation for small matrices with $m= \mathrm{BsizeD}_1 \times \mathrm{BsizeD}_2 \times \mathrm{BsizeD}_3$, $n=N_\mathrm{mom}$ and $k=\mathrm{BsizeX}$.
These matrices -- $singlet$ and \textit{tmpBuf} -- should be sufficiently small to remain in the cache.
General-purpose GEMM implementations are typically optimized targeting larger matrix sizes.
Thus, for the small matrix-matrix multiplication required here, the \mkl{} with JIT compilation is used to generate target microarchitecture code for the kernel which is optimized for small-sized complex-valued matrix multiplication problems. 
As the matrix sizes are fixed, the kernel can be produced once and then called many times, amortizing the JIT compilation overhead.

Parallelization for multiple threads is achieved by distributing the work in the loops over dilution-index blocks, (ll.~4-5) .
In practice, the loops over $\mathrm{BlockD}_1$ and $\mathrm{BlockD}_2$ are collapsed into a joint iteration space and parallelized for multiple threads with OpenMP.
In this approach, there are no data dependencies since computations of each thread are fully independent.
Furthermore, as each block works on an identical amount of data, the load is expected to be well-balanced between threads.
Lastly, parallelization is implemented at an outer level, encompassing plenty of work per loop trip to amortize the OpenMP runtime overhead for instance for thread scheduling.

%----------------------------------------------------------------------
\section{Performance results}
\label{sec:results}
\begin{figure}[tbh]
    \centering
    \includegraphics[width=0.48\textwidth]{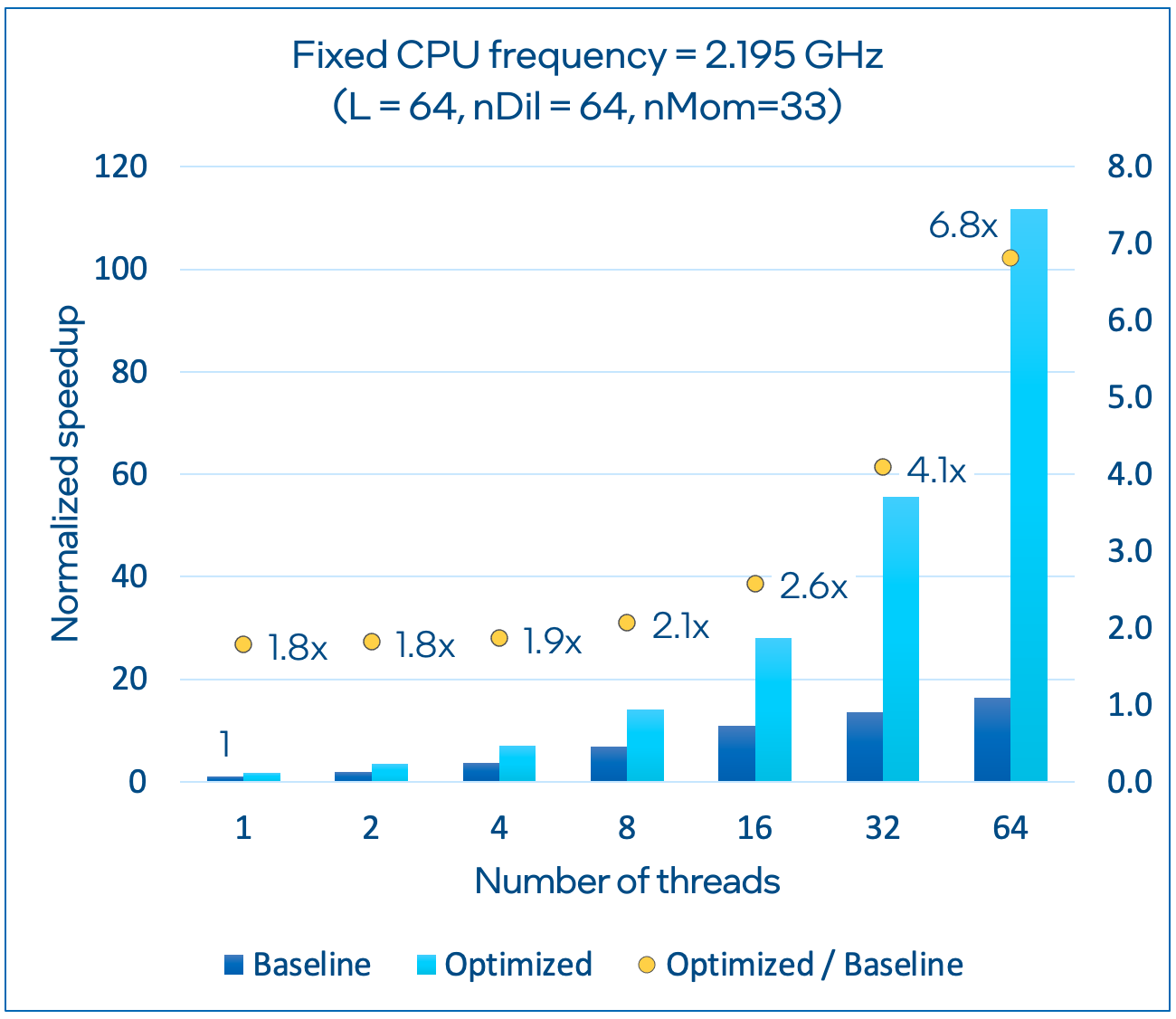}
    \includegraphics[width=.48\textwidth]{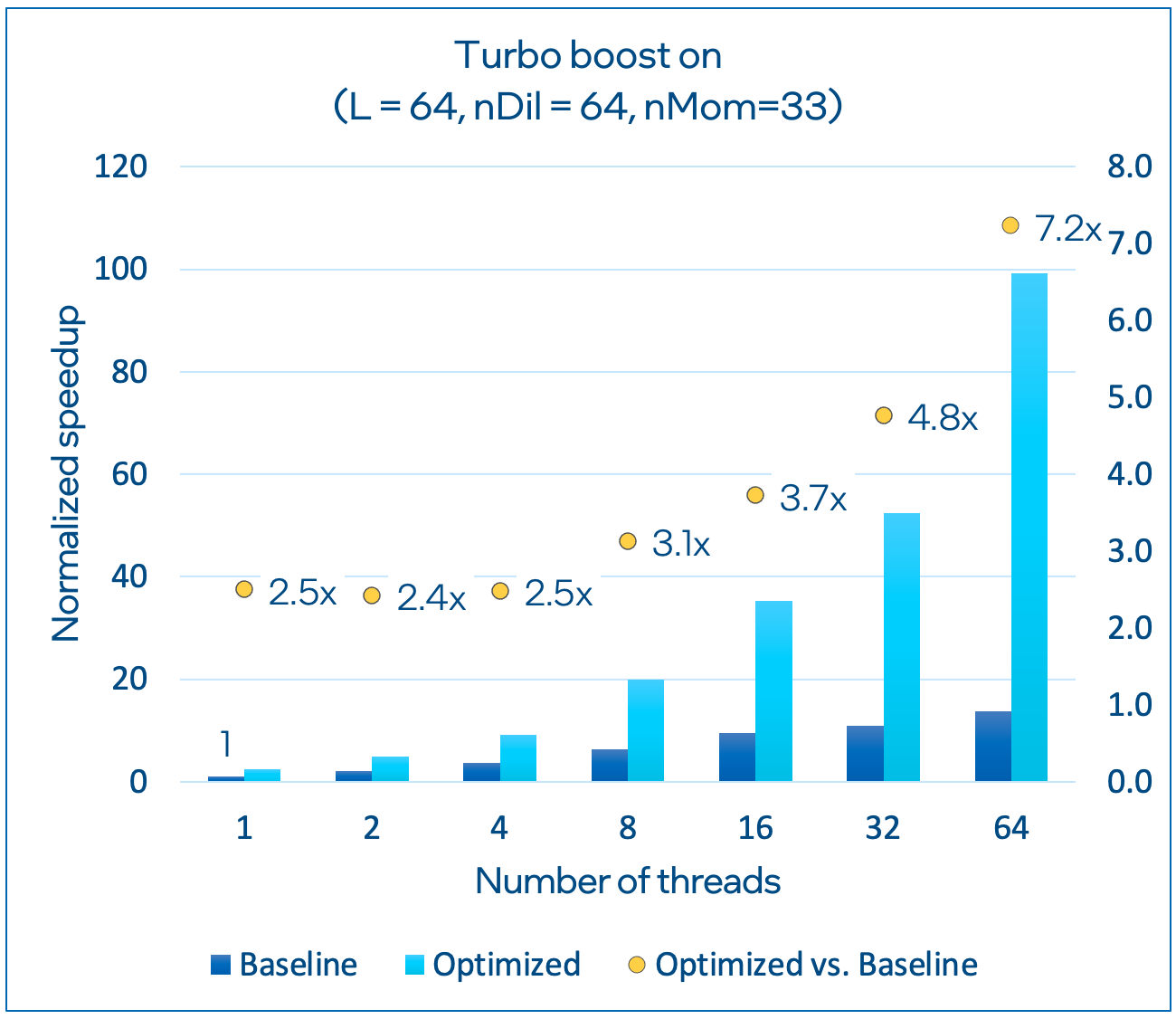}
    \caption{Performance of the previous and optimized kernel normalized to the single-thread performance of the previous implementation.
    The optimized kernel outperforms the baseline by up to 6.8x for the test with fixed frequency (left) and by up to 7.2x for the test when turbo boost is enabled (right).}
    \label{figs:cpus_compare}
\end{figure}

The implementation of Algorithm~\ref{alg:opt} is evaluated on a test system  with two Intel\SymbReg~Xeon\SymbReg~Platinum 8358 processors @ 2.60 GHz for a moderately large problem size of $L=64$ with $N_\mathrm{dil}=64$ dilution indices per quark field and number of requested momenta $n_\mathrm{mom}=33$. 
Figure~\ref{figs:cpus_compare} shows the performance of the optimized kernel compared to the previous implementation using 1 to 64 cores.

The block sizes are tunable parameters which generally depend on the target architecture and should be tuned experimentally. 
For our test node, the block sizes yielding the best performance are $\mathrm{BsizeX} = 32$, $\mathrm{BsizeD}_1 = 4$, $\mathrm{BsizeD}_2=8$, $\mathrm{BsizeD}_3=16$.  
This can be understood as aligning the memory footprint of each single intermediate object with the available cache hierarchy. 
For instance, the $diq$ array stores $\mathrm{BsizeD}_1 \times  \mathrm{BsizeD}_2 \times  N_\mathrm{colors} \times \mathrm{BsizeX} = 3072$ complex double-precision values, occupying 48 kB of memory, thus fitting perfectly into the L1 cache.
Similarly, the $singlet$ has a size of 256 kB which fits into the L2 cache.

When run on a single core, the optimized kernel is 1.8x and 2.5x faster than the baseline in tests with and without turbo boost, respectively.
With in-depth profiling, the superior performance of the optimized implementation can be traced back to better use of the memory system, as expected.
The optimized kernel operates at 1.6x and 3x higher arithmetic intensity in Read and Write, respectively.
In addition, the memory access patterns are also significantly improved such that fewer cycles are spent on load and store operations in the optimized kernel in comparison to the baseline (42\% reduction in loads and 70\% reduction in stores). 
The optimized kernel is indicated to be L1-bound instead of DRAM-bound.
As a result, with a performance of 37.7 DP GFlops/s, the kernel reaches 54\% of the theoretical peak performance. 
\begin{figure}[tbh]
    \centering
    \includegraphics[width=0.48\textwidth]{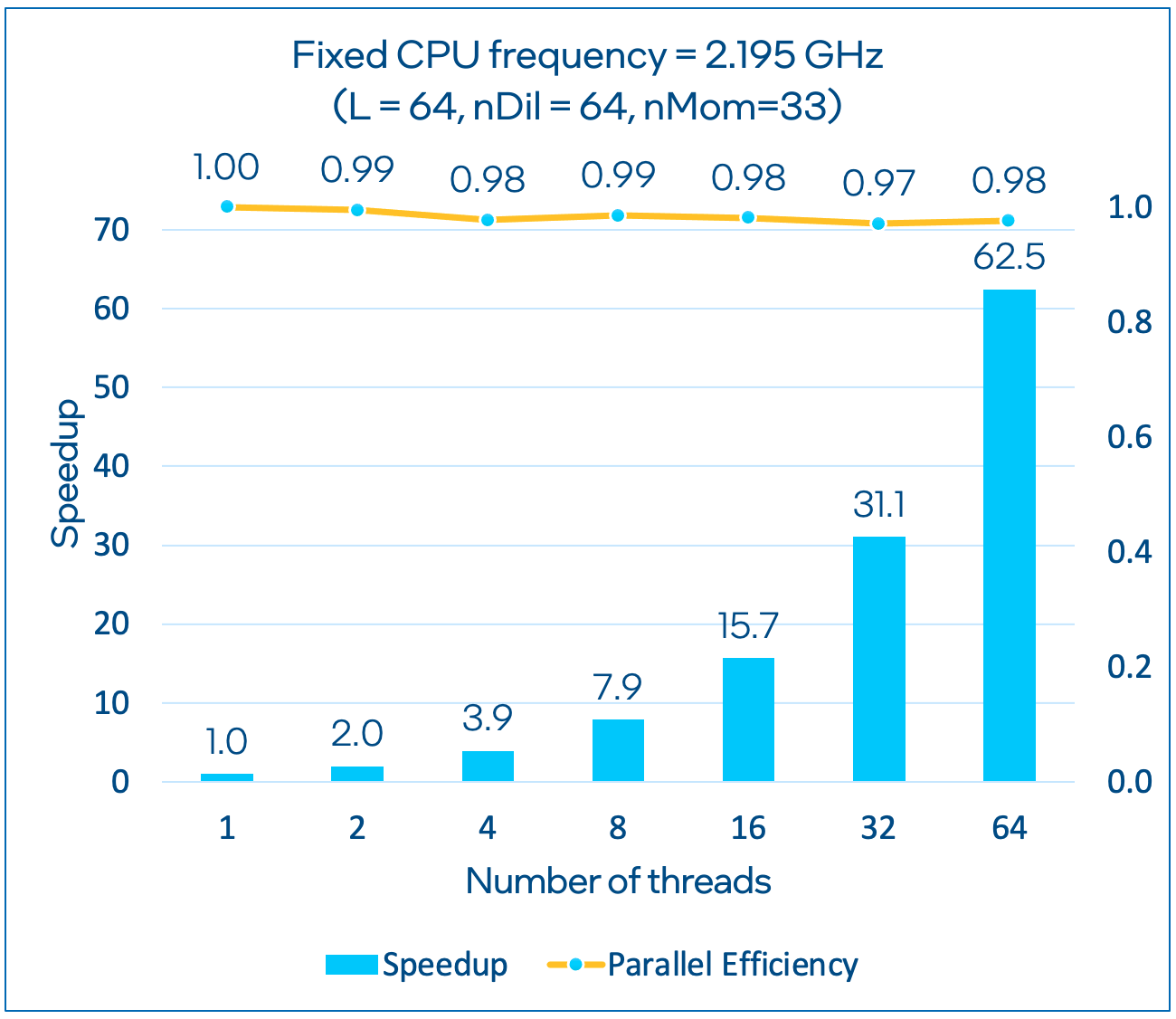}
    \includegraphics[width=.48\textwidth]{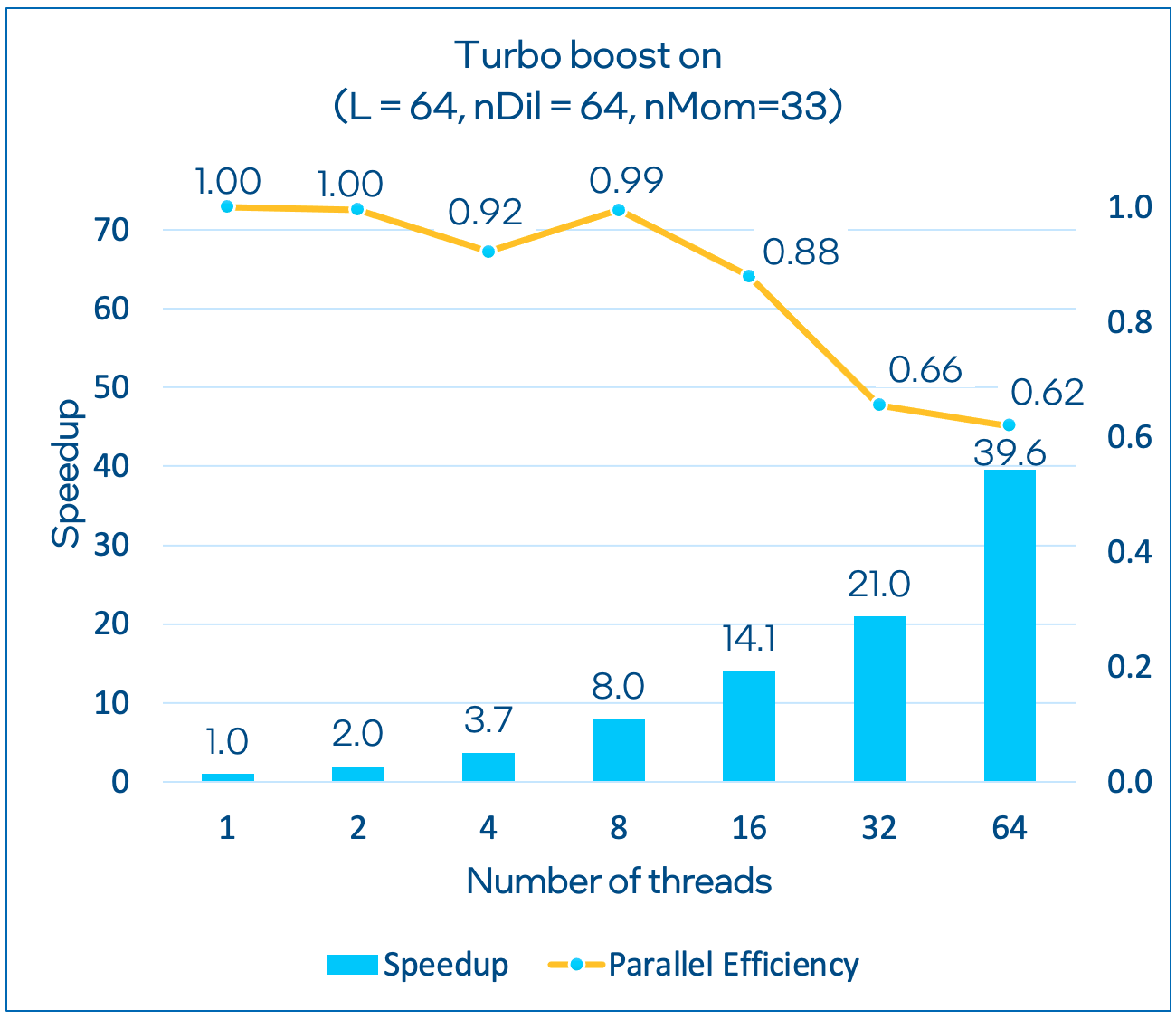}
    \caption{Strong-scaling behavior of the optimized kernel.
	    \textit{Left:} With fixed frequency, the kernel achieves almost perfect parallel efficiency for a 62.5x speedup with 64 threads.
	    \textit{Right:} With turbo enabled, the kernel scales well up to eight cores. Beyond eight cores, the variable frequency gets throttled and the decrease in parallel efficiency matches the decrease in frequency, implying that the optimized kernel is compute-bound.
    }
    \label{figs:cpus_strong}
\end{figure}

The improvement of the optimized kernel becomes particularly apparent in multithreaded runs,
where it outperforms the baseline by 1.8x to 6.8x at fixed frequency and 2.5x to 7.2x for tests with turbo boost on (\autoref{figs:cpus_compare}).
This boost in performance can be understood as being due to improved temporal locality.
As long as threads progress at a comparable rate, adjacent threads working on a different BlockD2 but sharing the same BlockD1 access the same input data, which may be served from the cache hierarchy.

Figure~\ref{figs:cpus_strong} shows the strong-scaling behavior of the optimized kernel at fixed clock frequency, achieving almost perfect parallel efficiency at a speedup of 62.5x when using 64 cores.
In addition, the parallel efficiency is above 0.98 for all tests, implying that the kernel scales almost perfectly within a node.

For production runs with turbo boost enabled the clock frequency can increase up to the boost frequency (3.6 GHz on our test system).
While performance in absolute terms is slightly better than at fixed frequency, the strong-scaling parallel efficiency deteriorates, showing a 39.6x speedup with 64 cores (\autoref{figs:cpus_strong}).
The loss in parallel efficiency is due to frequency throttling. While for one thread the average frequency is $3.285 \, \mathrm{GHz}$, it drops to $2.219 \, \mathrm{GHz}$ when running with 32 threads.
The frequency ratio $0.67$ matches the parallel efficiency, indicating that the optimized kernel is compute-bound and indeed limited by the frequency throttling.

\begin{figure}[tbh]
    \centering
    \includegraphics[width=0.48\textwidth]{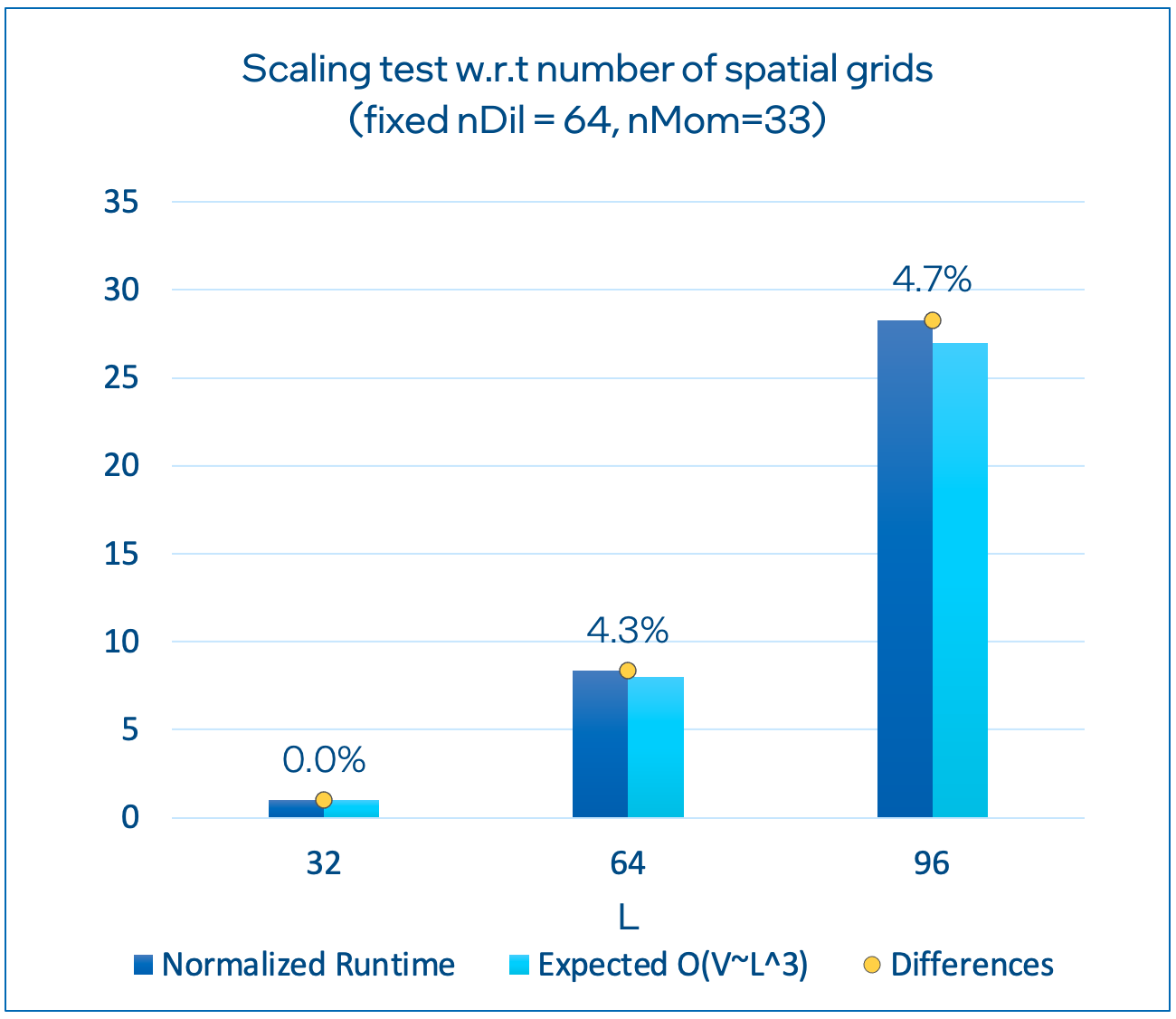}
    \includegraphics[width=.48\textwidth]{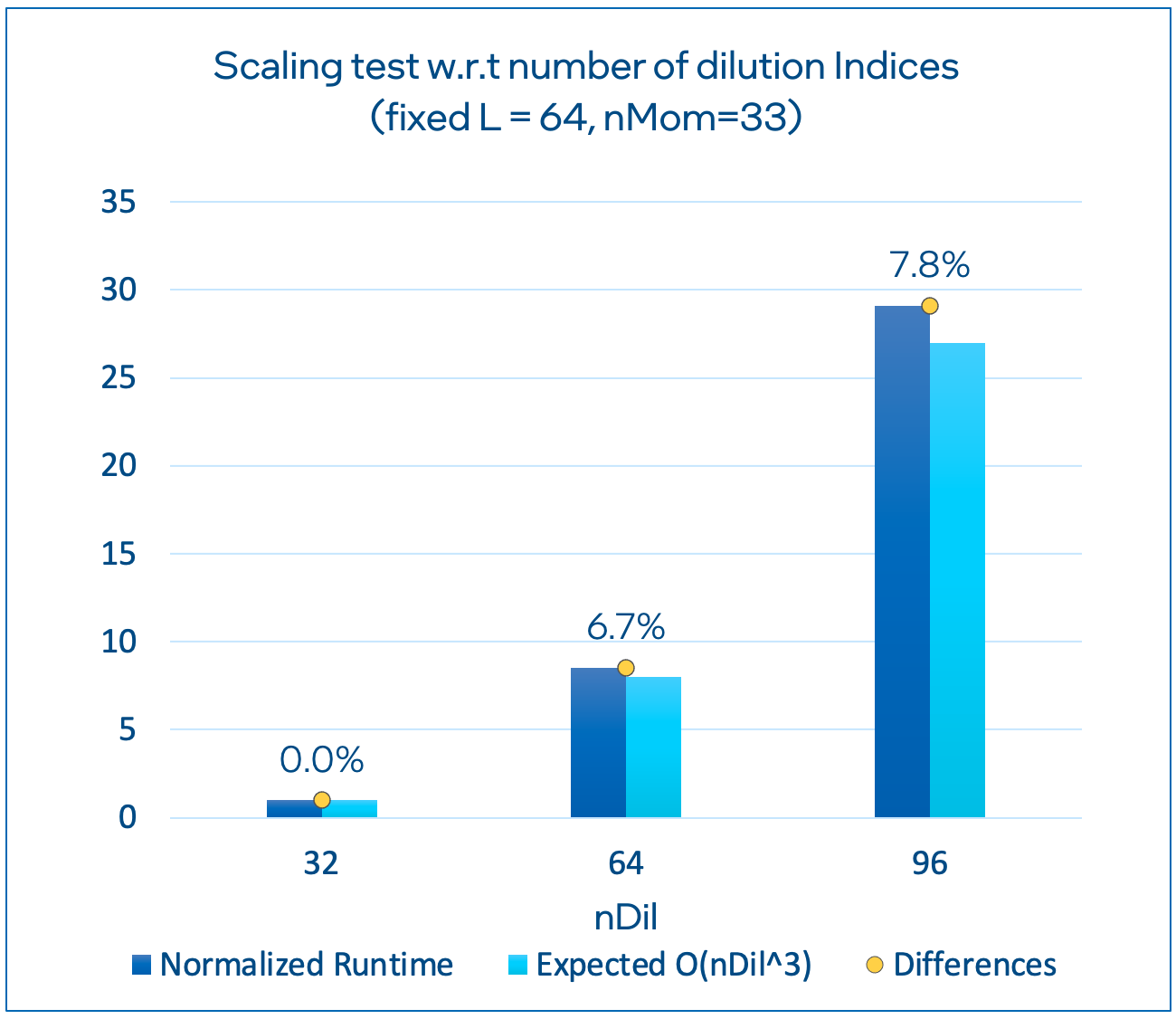}
    \caption{Scaling performance of the optimized kernel with respect to the problem size. The kernel scales as expected to within a few percent both with the spatial volume and the number of dilution indices.}
    \label{figs:cpus_twoscale}
\end{figure}
The scalability of the optimized kernel with respect to the problem size, of importance in view of ever-increasing simulation volumes, is shown in Figure~\ref{figs:cpus_twoscale}.
As is evident from \eqref{eqn:bkern}, the computational cost scales as $\mathcal{O}(L^3)$ and $\mathcal{O}(N_\mathrm{dil}^3)$ with the spatial volume and number of dilution indices, respectively.
The optimized kernel scales as expected to within a few percent as a function of the problem size both with the spatial volume as well as the number of dilution indices.

%--------------------------------------------------------------------------
\section{Summary}
We have presented an optimized implementation of the kernel computing baryon blocks in the stochastic LapH method, achieving an up to 7.2x speedup over the previos implementation.
Exploiting the high arithmetic intensity of \eqref{eqn:bkern} by blocking in dilution and spatial indices, and performing the momentum projection with a JIT-compiled microkernel provided by the \mkl{}, we achieve good single-core performance on a test system with Intel\SymbReg~Xeon\SymbReg~Platinum 8358 processors.
Parallelizing over blocks of dilution indices using multithreading, we also observe good scalability all the way to the maximum number of cores per socket%
\footnote{Depending on memory requirements, this parallelization can straightforwardly be supplemented by a standard domain decomposition over MPI ranks in order to enable scaling beyond one socket or node at the expense of an $N_\mathrm{dil}^3$-sized MPI reduction at the end of kernel execution}.

This optimized implementation has been upstreamed into the \texttt{chroma\_laph} measurement suite and is ready for use in production runs on large lattice volumes.

\acknowledgments
We acknowledge useful conversations with Colin Morningstar as well as our Intel colleagues Christoph Bauinger and Carsten Uphoff. Intel and the Intel logo are trademarks of Intel Corporation or its subsidiaries.

\bibliographystyle{JHEP}
\bibliography{refs}
\end{document}